# Direct Comparison of Isobaric and Isochoric Vitrification of Two Aqueous Solutions with Photon Counting X-Ray Computed Tomography

Jason T. Parker[*][†], Anthony N. Consiglio[*][§], Boris Rubinsky, Simo A. Mäkiharju

*Department of Mechanical Engineering, University of California, Berkeley, Berkeley, California 94720, USA*

[*]These authors contributed equally.
[†]jtparker@berkeley.edu
[§]aconsiglio4@berkeley.edu

## Abstract

Vitrification is a promising approach for ice-free cryopreservation of biological material, but progress is hindered by the limited set of experimental tools for studying processes in the interior of the vitrified matter. Isochoric cryopreservation chambers are often metallic, and their opacity prevents direct visual observation. In this study, we introduce photon counting X-ray computed tomography (CT) to compare the effects of rigid isochoric and unconfined isobaric conditions on vitrification and ice formation during cooling of two aqueous solutions: 50 wt% DMSO and a coral vitrification solution, CVS1. Previous studies have only compared vitrification in isochoric systems with isobaric systems that have an exposed air-liquid interface. We use a movable piston to replicate the surface and thermal boundary conditions of the isochoric system yet maintain isobaric conditions. When controlling for the boundary conditions we find that similar ice and vapor volume fractions form during cooling in isochoric and isobaric conditions. Interestingly, we observe distinct ice and vapor cavity morphology in the isochoric systems, possibly due to vapor outgassing or cavitation as rapid cooling causes the pressure to drop in the confined systems. These observations highlight the array of thermo-fluidic processes that occur during vitrification in confined aqueous systems and motivate the further application of imaging techniques such as photon counting X-ray CT in fundamental studies of vitrification.

## 1. Introduction

The ability to preserve biological systems such as cells, plants, organs, and even entire organisms for indefinite periods of time has remained an elusive goal of the biomedical and conservation communities as traditional clinical methods only enable the storage of organs for a few hours [14] and cryo-banking of full organisms is not yet possible. The significant challenge of preserving biological systems outside of their normal environment stems from the requirement of either 1) maintaining normal physiological processes that support metabolic activity [5,17] or 2) suppressing metabolic processes to prevent deterioration. Metabolism, which represents the sum of enzymatic reactions that sustain life, is largely an Arrhenius process, and as such, is a strong function of temperature. At cryogenic temperatures, molecular processes are arrested to such an extent as to bring about a state of suspended animation.

The greatest challenge facing successful cryopreservation is avoiding the formation of ice, which is nearly universally lethal to biological systems through various mechanical, chemical, and osmotic mechanisms [23]. Vitrification involves the solidification of a liquid not by crystallization but rather through formation of an amorphous glass whereby an increase in viscosity results in arrested molecular motion and suppressed ice nucleation. Vitrification therefore represents a compelling method for ice-free cryopreservation. However, the nucleation rate of ice in aqueous solutions is a steep function of supercooling, and cooling rates exceeding $10^6$ °C/s are required to vitrify water [3,6]. The addition of solutes, termed cryoprotective agents (CPAs), and the application of pressure affect both equilibrium and non-equilibrium freezing processes and lower the cooling rate needed to achieve glass formation [12]. At the concentrations needed to effectively suppress ice nucleation many CPAs are toxic, further complicating the application of vitrification and motivating the search for new methods to achieve stable ice-free cryopreservation.

Recently, rigid isochoric confinement has been found to stabilize supercooled water against ice nucleation by providing isolation from external perturbations, removing exposed liquid-air interfaces, and potentially imparting an energetic toll upon incipient ice embryos as they expand relative to the liquid phase [9–11,25]. Confinement may also alter ice growth dynamics since the formation of ice in a confined volume acts to self-pressurize the system, which shifts the phase equilibrium and can result in arrested crystallization dynamics and multi-phase solid-liquid equilibrium [28]. In the context of this study, isochoric confinement refers to nominally rigid systems in which the pressure fluctuations are more significant than volume with changes of temperature. Conversely, isobaric refers to a roughly constant pressure system in which the volume is free to fluctuate with temperature.

Isochoric confinement has also received attention for its potential to improve cryopreservation outcomes by promoting vitrification. In an exploratory study, Zhang et al. [36] monitored pressure during cooling as an indicator for ice formation and found that concentrations lower than the critical values reported in the literature vitrified in isochoric systems without an increase in pressure. Solanki and Rabin subsequently proposed in a computational study that contraction of an aqueous solution during cooling would generate vapor cavities, enabling the absence of a pressure rise in the presence of certain ice fractions and potentially explaining the previous observations [31]. More recently, Powell-Palm et al. [24] employed an isochoric system in order to successfully vitrify and revive centimeter-scale coral specimens, representing an important cryo-conservation milestone. To the best knowledge of the authors, Powell-Palm et al. [24] are the only group to experimentally compare isochoric and isobaric vitrification of aqueous systems. That study compares an enclosed isochoric chamber to an open isobaric chamber with an air-liquid interface. As a result, the heat transfer and surface conditions are not matched between the isochoric and isobaric configuration. Comparing isochoric and isobaric vitrification with matched heat transfer and surface conditions has remained unexplored.

Motivated by the promising vitrification result of Powell-Palm et al. [15] and the still poorly understood process of vitrification in confined systems, we adapted our custom-built photon counting CT system [20, 21] to conduct photon counting X-ray computed tomography (CT) measurements of two aqueous solutions cooled at rates of roughly 100 °C/min in liquid nitrogen (50 wt% dimethyl sulfoxide (DMSO) and the coral vitrification solution (CVS1) [24]), each under isochoric and isobaric conditions. This technique uses a laboratory-scale X-ray system to resolve phase segregation inside rigid metallic chambers at $O(100 \ \mu m$ reconstructed resolution) and captures the full aqueous volume in a single scan.

We find that vitrification of 50 wt% DMSO leads to an inward tapered interface in both isochoric and isobaric configurations, which is a result of thermal contraction of the fluid. In the isochoric system the chamber pressure drops as the fluid contracts and a vapor cavity (or cavities) forms through either outgassing or cavitation. The CVS1 solution is found to partially crystallize in nearly equal amounts when cooled at 100-120 °C/min in both isochoric and isobaric conditions. Distinct ice morphologies are observed in CVS1 between the isochoric and isobaric-piston configurations, potentially resulting from similar cavity formation processes as in the 50 wt% DMSO solution – namely, outgassing or cavitation. These observations highlight the role of thermo-fluidic processes in the vitrification of confined aqueous systems and pave the way for future investigations of ice nucleation, growth, and vitrification processes using imaging techniques such as photon counting X-ray CT.

## 2. Materials and methods

### 2.1. Vitrification solutions

We selected two aqueous solutions as the focus of these exploratory experiments. The first solution is a common binary glass-forming aqueous solution consisting of 50 wt% DMSO in deionized water. This solution has a well-characterized and easily achievable critical cooling rate of $\leq 5$°C/min [18], which enables us to isolate the effects of isochoric confinement on solution contraction in the absence of ice formation as well as evaluate previous computational models of thermo-mechanical processes during vitrification [26,31,32]. The second solution, known as CVS1, was used in isochoric systems by Powell-Palm et al. [24] to cryopreserve whole coral specimens and is selected in order to provide insights into this promising cryopreservation result. The CVS1 solution consists of 1.05M DMSO, 1.05M propanediol, 1.05M glycerol, and 0.85M trehalose in physiological buffered saline (PBS) (see Supplemental Material for the ingredient weight fractions). Since CVS1 is comprised of multiple traditional CPAs as well as a disaccharide, the critical cooling rate is difficult to estimate *a priori*. Previous studies have not rigorously characterized the critical cooling rate of CVS1, but cooling at roughly 100°C/min was found to yield positive outcomes during the coral cryopreservation [24]. The experiments presented herein aim to investigate vitrification processes in this newly employed solution.

## 2.2. Vitrification chambers

Figure 1 depicts the three systems employed in this study. The main chamber body, constructed of aluminum-7075, has an inner diameter of 12.7 mm (0.5"), an outer diameter of 34.9 mm (1-3/8"), and a nominal internal volume of 5.3 ml. Multiple chambers are used throughout the experiments and all have either anodized or naturally oxidized surfaces. An aluminum threaded plug with a metal-on-metal tapered sealing surface provides rigid confinement for the isochoric trials, as shown on the left in Figure 1. Due to the thermal contraction of aluminum, the internal chamber volume contracts by approximately 1.5% during cooling in liquid nitrogen. We measure the volume of the chamber after contraction using photon counting X-ray CT. In order to study the solutions under unconfined isobaric conditions, we introduce two additional configurations. In the first configuration (Figure 1 right), the liquid interface is open to air and the chamber is covered with Kapton tape to prevent the intrusion of liquid nitrogen during cooling. This configuration allows the liquid interface to freely deform and is similar to systems modeled in previous simulations of thermo-mechanical processes during vitrification [26,32]. In trials with CVS1 in this configuration, we observed that slower cooling rates near the top surface of the fluid, which is not directly in contact with the metallic chamber, resulted in significant ice formation (see Figure 6), growing vertically out of the field of view of the X-ray detector. This observation and our desire to examine effect of matching thermal and surface boundary conditions of the isochoric plug motivated the introduction of the second isobaric configuration (Figure 1 center), which employs a vertically unconstrained aluminum piston. To ensure that no air is trapped underneath the piston, the face of the piston (or plug in the isochoric configuration) is wetted with the solution prior to insertion into the chamber. Solution is drawn out from the TC hole (used as weep hole) with a syringe after the plug is inserted. A preliminary X-ray image is captured prior to cooling to confirm that no bubbles are present.

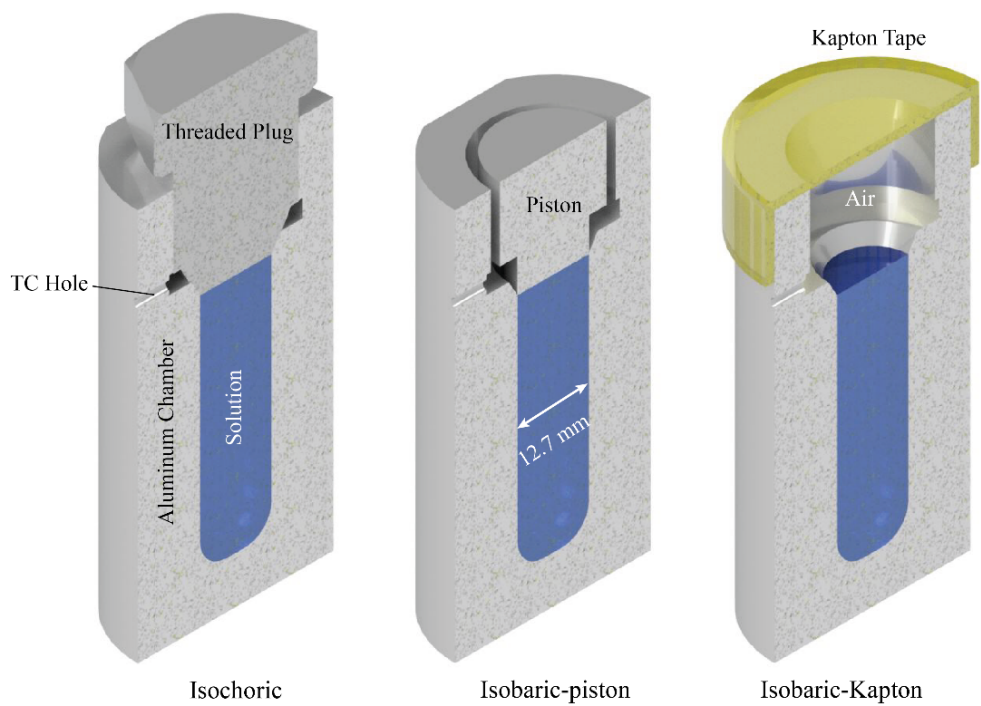

**Figure 1: Vitrification chambers.** A two-piece assembly with metal-on-metal sealing surfaces generates rigid isochoric confinement (left). Replacing the rigid plug with a movable aluminum piston (center) produces isobaric conditions while resulting in nominally equivalent thermal and surface conditions as the isochoric system. This configuration is employed in the trials with CVS1. Critical cooling rates of ≤ 5 °C/min are easily achievable with 50 wt% DMSO, and the solution is left uncovered in order to allow the free interface to deform. In this configuration (right), the chamber is capped with Kapton tape to prevent intrusion of liquid nitrogen during cooling. Temperature is monitored using a type-T thermocouple inserted partially into the side of the chamber (TC hole).

**Table 1:** Average cooling rates (°C/min) during plunge cooling in liquid nitrogen of CVS1 and 50 wt% DMSO under isochoric and isobaric conditions. Chamber configurations are depicted in Figure 1.

|  | CVS1 | | | 50 wt% DMSO | |
|---|---|---|---|---|---|
|  | Isochoric | Isobaric (piston) | Isobaric (Kapton) | Isochoric | Isobaric (Kapton) |
| Trial 1 | 120 | 108 | 119 | 112 | 109 |
| Trial 2 | 106 | 117 | 134 | 116 | 120 |
| Trial 3 | 100 | 104 | 114 | 110 | 127 |
| Average | 110 | 109 | 122 | 113 | 119 |

### 2.3. Vitrification procedure

In order to facilitate rapid cooling, we lower a chamber in a vertical orientation $O$(10 cm/s) to the bottom of a 20-30cm deep liquid nitrogen bath. The temperature is monitored during this process with a type-T thermocouple inserted into the wall of the chamber, as indicated in Figure 1, and recorded with an Omega TC-08 thermocouple data logger (with a manufacturer specified uncertainty of ±1 °C). Due to the size and thermal mass of the systems, thermal gradients are present during the plunge cooling, and we utilize this temperature measurement as a rough indicator of a chamber's thermal history. The average cooling rate is computed for each trial from the recorded temperature measurement and is reported in Table 1. We conduct three repetitions for each condition to account for potential variability in assembly condition and cooling profile.

### 2.4. Photon counting X-ray computed tomography

Figure 2 depicts the X-ray CT setup, which consists of a YXLON FXE225.48 X-ray source used in concert with a Dectris Pilatus3 100K-M photon counting detector. Photon counting detectors (PCDs) achieve a higher signal-to-noise ratio than conventional X-ray detectors [29], making them advantageous for collecting CT scans of systems with low X-ray contrast such as multiphase vitrified systems. Lower energy photons, which are more likely the result of scatter and thus introduce noise, can also be rejected with PCDs. That PCDs are photon energy-resolving detectors instead of the traditional photon energy-integrating detectors is the primary advantage and the root of their benefits. The energy threshold for these experiments is set at a nominal 20 keV with a 10 keV threshold roll-off setting on the Pilatus3 detector. We collect 114 projections per CT scan, each with a 9999.05 ms integration time and a 0.95 ms dead time for a total imaging period of 10 s per projection. The rotation rate is constant at 0.316 degrees per second.

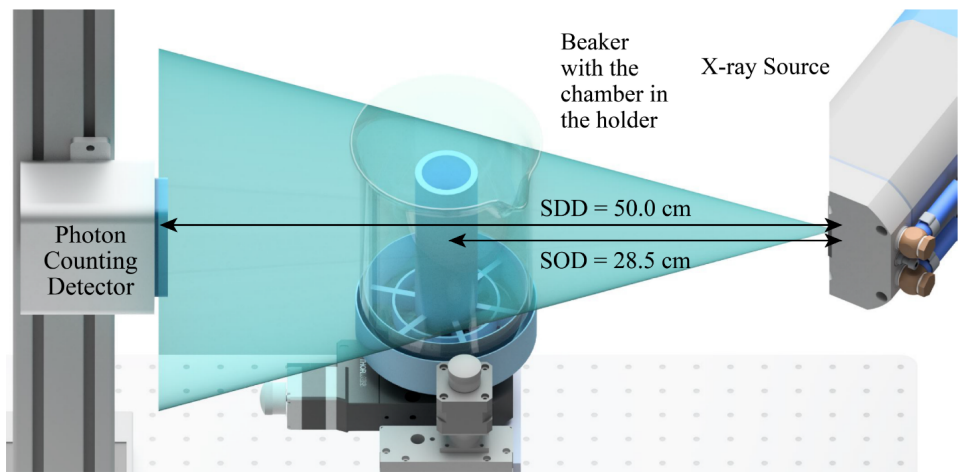

**Figure 2:** Experiment setup consisting of a cone beam X-ray source and a photon counting detector. After cooling in a separate dewar of liquid nitrogen, the chamber is placed into a beaker

that sits atop a stack of three linear stages and one rotation stage. The beaker is partially filled with liquid nitrogen below the chamber to ensure the chamber does not warm over the duration of the CT scan but without obstructing the main field of view.

The X-ray source emits a polychromatic cone beam, which produces geometric magnification based on the positions of the object and the detector relative to the source. The magnification factor, M, is the ratio between the source-to-detector distance (SDD) and the source-to-object distance (SOD). Based on a magnification factor of M = 1.75 and a pixel size of 172 μm square, we achieve a spatial resolution of 98 μm square for a single projection. Gaining higher spatial resolution sacrifices the field of view, however, reducing the nominal detector area of 8.38 cm high by 3.35 cm long to an effective field of view measuring 4.79 cm high by 1.91 cm long. Although a 98 μm square projection spatial resolution is achieved, the true reconstructed spatial resolution is likely lower due to the relatively few projections captured. The reconstructed resolution is $O(100 \ \mu m)$.

We operate the X-ray source at an acceleration voltage of 80 kVp and a target current of 0.8 mA. Based on the manufacturer's specifications, the focal spot for the source at these settings is approximately 87.5 μm full-width half-max. With geometric magnification, we expect the focal blurring to be approximately 153 μm. This is less than the size of a single pixel, so we do not expect notable focal blurring artifacts. Further details on the X-ray imaging setup we use can be found in [21,22].

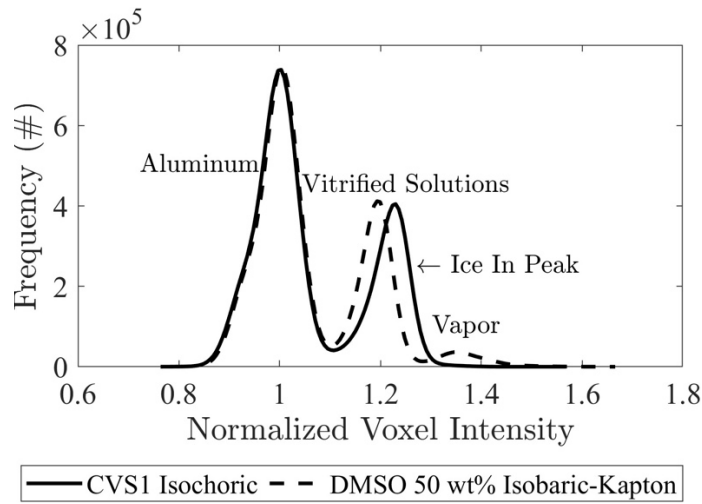

**Figure 3:** Normalized voxel intensity histograms of the CT scan reconstruction for the entire volume of CVS1 and 50 wt% DMSO rapidly cooled in liquid nitrogen in isochoric and isobaric-Kapton configurations, respectively. These histograms illustrate the contrast between the aluminum chamber, the vitrified solution, and the vapor phases. The histograms are aligned based on the aluminum peaks.

## 2.5. Phase identification

We use the filtered back projection algorithm in the ASTRA Toolbox [1,2] to reconstruct the CT scans onto a 195 x 195 x 487 grid with $(98 \ \mu m)^3$ voxels. A normalization ring artifact correction is applied with a standard deviation of 10 pixels. We then segment the phases in ORS Dragonfly [19] based on their voxel intensity in order to calculate their respective volumes and assess morphology. A 3D median filter with a span of 5 voxels is used to enhance the contrast between ice and the vitrified solution. We do not median filter the vapor cavity scans because there is already sufficient contrast. Figure 3 depicts the normalized voxel intensity histograms for the entire reconstructed volume of CVS1 and 50 wt% DMSO under isochoric and isobaric conditions, respectively. These images demonstrate the contrast between ice, vapor, and vitrified solution with multiple peaks visible. Starting from the left: the first is the aluminum chamber walls; the second is the vitrified solution and, in the CVS1 solution, ice; the third is a vapor cavity, which is only present in the 50 wt% DMSO solution. The difference in the 50 wt% DMSO and CVS1 vitrified solution peaks is due to the higher X-ray attenuation of the 50 wt% DMSO solution compared to CVS1 across the range of photon energies (see Supplemental Material figure S1). In this depiction, the ice voxels in the CVS1 solution are hidden within the vitrified solution peak. Figure 4 depicts the histogram of a single unfiltered slice that contains ice in a CVS1 isochoric scan, revealing the distinction between ice and the surrounding vitrified solution. Each phase is segmented from the others by establishing thresholds at the local minimum between peaks.

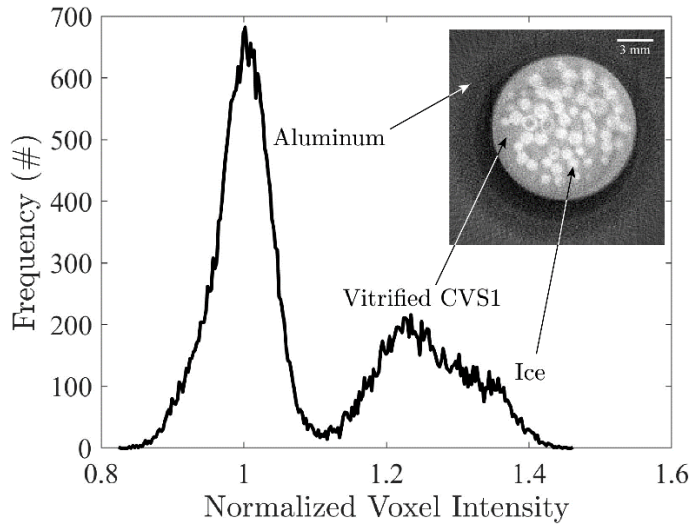

**Figure 4:** Voxel intensity histogram of a CT scan reconstruction for a single unfiltered slice of CVS1 rapidly cooled under isochoric conditions. Peaks are visible for the three present phases: aluminum, vitrified CVS1, and ice. Inset depicts the unfiltered reconstructed slice with globular ice morphology.

## 3. Results and discussion

### 3.1. Vitrification of 50 wt% DMSO

The 50 wt% DMSO solution vitrifies completely with cooling rates exceeding 5 °C/min and is selected for investigating thermal contraction behavior. Trials with this solution in the isobaric-piston configuration were not conducted because the slow critical cooling rate obviates the need for heat transfer optimization. Figure 5 depicts representative 3D reconstructions of the vitrified 50 wt% DMSO solution in isochoric and isobaric-Kapton configurations. All reconstructions are available in the Supplemental Material figures S2 and S3. A contracting taper morphology is observed in both conditions and is very similar to deformations observed in previous computational studies of vitrifying systems with free surfaces [26,32]. Average vapor volume fractions of 5.7% and 6.7% are computed for the isochoric and isobaric conditions, respectively. Table 2 reports the volume fractions of vapor from each trial. As predicted by the recent computational study by Solanki and Rabin [31], vapor cavities are observed to form during cooling of DMSO solutions in the isochoric chamber. The solution contracts more than the chamber, causing the pressure to drop below the saturation pressure and results in the rejection of dissolved gasses or, if the pressure drops quickly enough, vapor cavitation. The formation of a vapor cavity nullifies the original isochoric confinement.

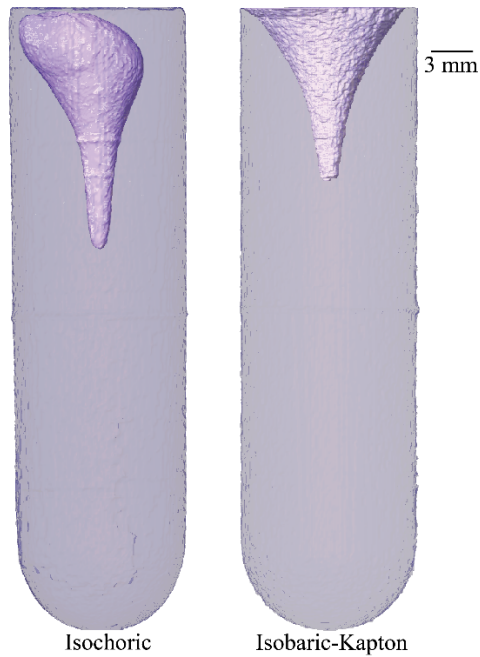

Isochoric       Isobaric-Kapton

**Figure 5:** Reconstructed CT scans of 50 wt% DMSO vitrified under isochoric (left) and isobaric-Kapton (right) conditions. Solution contraction leads to surface deformation and, in the isochoric system, vapor cavity formation. Vitrified solution and vapor phases are segmented and shown in purple and white, respectively. While the vitrified solution is pinned around the brim of the chamber in the isobaric system, a film of vitrified solution covers the entire interior surface of the isochoric system. Vapor volume fractions are reported in Table 2.

Interestingly, while the vitrified solution is pinned at the brim of the chamber in the isobaric system, a thin film of vitrified solution is found to remain over the entire inner surface of the isochoric system. The sharp interface in the isobaric system may serve as a local stress concentration and could be susceptible to fracturing upon rapid rewarming [26]. Such stresses may be ameliorated by the smooth interfaces in the isochoric system. The current study only considers static equilibrium systems at cryogenic temperatures; future studies may expand on this and look into how these differences in vitreous interface formation affect glass stability upon rewarming in isobaric and isochoric systems.

### 3.2. Ice formation during vitrification of CVS1

Due to its lower molarity, CVS1 has a higher critical cooling rate and a lower coefficient of thermal expansion than the 50 wt% DMSO solution. It is therefore well suited for investigating the potential for isochoric confinement to promote vitrification. The CVS1 solution is cooled following the protocol employed in Powell-Palm et al. [24] resulting in average cooling rates of about 110°C/min (see Table 1). Despite this protocol enabling cryopreservation and revival of coral specimens (in the isochoric configuration), we find that a small amount of ice crystallizes during cooling.

Figure 6 shows representative 3D reconstructions of the ice formed in CVS1 in the isobaric-piston, isochoric, and isobaric-Kapton systems. All reconstructions are available in the Supplemental Material figures S4 – S6. As discussed previously, a significant amount of ice forms in the isobaric-Kapton configuration because the exposed liquid interface does not receive the same direct cooling from the highly conductive aluminum chamber. This observation motivated the introduction of a movable piston to replicate the thermal boundary condition from the isochoric system while maintaining isobaric conditions. When controlling for the cooling profile in this way, we find that isochoric and isobaric-piston conditions yield similar amounts of ice (on average 0.56% and 0.78%, respectively). The volume fractions of resolvable ice formed in each trial are reported in Table 2. It is worth noting that the volume fractions are computed based on resolvable phase segregation at a reconstructed resolution $O(100 \ \mu m)$. The nucleation rate of ice is known to reach a maximum at temperatures close to the glass transition temperature while the rate of ice crystal growth rapidly slows at much higher temperatures. During vitrification, this results in the potential for sub-micron crystals that nucleate at very low temperatures but cannot grow to an appreciable size due to the slowed growth kinetics [33]. Future efforts should seek to increase the spatial resolution in order to resolve smaller ice crystals.

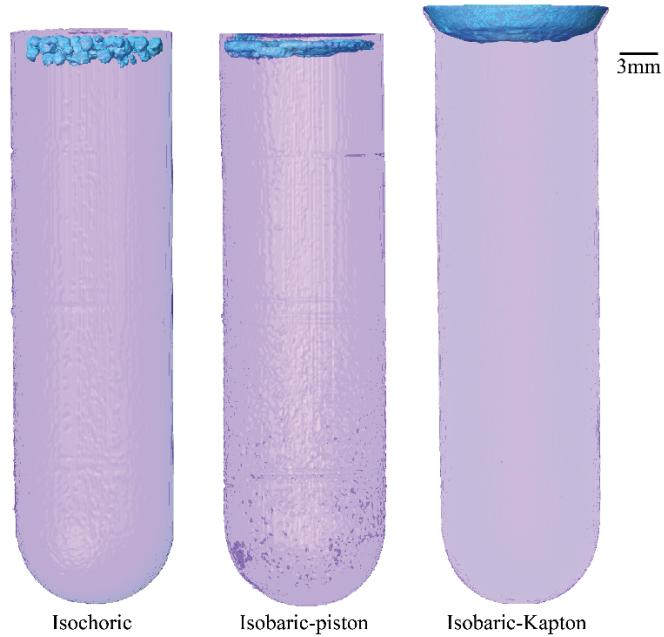

**Figure 6:** Reconstructed CT scans of CVS1 in isochoric (left), isobaric-piston (center), and isobaric-Kapton (right) configurations. The vitrified solution and ice phases are segmented and shown in purple and blue, respectively. While a single mass of ice forms in the isobaric-piston system, many millimeter-scale ice globules form in the isochoric system. Slow cooling near the free interface in the isobaric-Kapton system results in significant ice formation that grows out of the measurement field of view. Ice volume fractions are reported in Table 2.

Despite finding similar ice volume fractions in CVS1 between the isochoric and isobaric-piston configurations, we do observe morphological differences in the ice that forms. Figure 7 depicts close-up reconstructions of the segmented ice phases in CVS1 for both conditions. The ice forms near the surface at the gravitational top of the chamber in both conditions. We conducted one experiment with an inverted isochoric system and found that the ice still formed at the gravitational top, suggesting that the observed ice is not a result of geometrical differences between the round bottom and flat lid. Whereas the ice in the isobaric system formed one continuous mass, the ice in the isochoric system formed many millimeter-scale globular crystals. This difference could be the result of outgassing or cavitation [4,13]. Under isochoric confinement the pressure in the chamber drops as the solution contracts during cooling. As the pressure drops below the equilibrium vapor pressure dissolved gasses may be forced out of solution. If the pressure drops rapidly enough, the solution may cavitate. Both possibilities are noted by Solanki and Rabin [31]. Vapor bubbles may then serve as a path of least resistance to ice growth, potentially resulting in the observed morphology.

The observation of ice formation in CVS1 under isochoric conditions also indicates that the critical cooling rate of CVS1 may be slightly higher than suggested by the experiments of Powell-Palm et al. [24]. That study estimated the critical cooling rate to be approximately 100 °C/min, as that cooling rate resulted in no pressure increase while slightly slower cooling rates did. Our observations of ice formation in CVS1 indicate that pressure measurements may not be a sufficient indicator of full vitrification on its own. A rise in pressure rise indicates ice formation, but a lack of a pressure rise does not preclude ice formation. The vapor cavities in the 50 wt% DMSO solution suggest that ice could displace the vapor before generating a detectable pressure rise, as proposed by Solanki and Rabin [14]. Nevertheless, for volumes of ice large enough to generate detectable pressure, pressure measurements represent a simple and fast method for probing the formation and relative quantity of ice formed during cooling based on the magnitude of the pressure rise. The relationship between ice fraction and pressure in isochoric systems depends on multiple factors: the solution thermal contraction and compressibility; the specific volume of the system, which itself depends on conditions of the assembly; and lastly the thermal contraction and compressibility of the chamber material. In the case of a highly contractile solution such as 50 wt% DMSO, up to approximately 6% of the system may crystallize before an increase in pressure is detected based on the observed thermal contraction. In less contractile solutions such as CVS1, detection of much smaller volumes of ice may be possible. Future studies may seek to characterize the thermophysical properties of vitrification solutions in order to better understand volumetric and pressure responses during vitrification in isochoric systems.

### 3.3. Cavity formation in isochorically confined aqueous solutions

The observed ice morphology in the CVS1 solution suggests that fluid contraction during cooling may produce cavity formation. Determining the occurrence of cavitation would require up to $O(1 \text{ MHz})$ frame rate measurements. Alas, isochoric vitrification occurs in a thick-walled metal chamber, necessitating relatively long exposure periods with most in-lab X-ray sources. Cryogenic temperatures in a boiling nitrogen bath additionally hinder the use of acoustic inspection or accelerometers. High-speed synchrotron X-ray imaging could potentially be well suited for determining the presence of cavitation during vitrification in isochorically confined systems.

If outgassing occurs, the morphology of rising bubbles in a fluid is a well understood function of the dimensionless Reynolds, Eotvos, and Morton numbers [8]. However, these parameters are difficult to estimate during vitrification as the kinematic viscosity of the fluid and the surface tension of the gas change rapidly with temperature and pressure. As a result, analytically or numerically confirming outgassing is challenging. Observing outgassing experimentally may be possible with slower frame rate in-lab imaging if the maximum outgassing rates are sufficiently slow. Should that be the case, it would be possible to rule out or confirm the presence of outgassing. We do not perform this step here as size constraints preclude us from placing the liquid nitrogen dewar in front of the X-ray source. As in the 50 wt% DMSO solution, the formation of vapor cavities would nullify the original isochoric confinement for the duration of their presence. Isochoric confinement would be restored should ice completely displace the vapor, for example. The presence of vapor does not imply that the system would necessarily be isobaric, however. During cavitation the pressure could change dramatically throughout the chamber, especially near the vapor cavity.

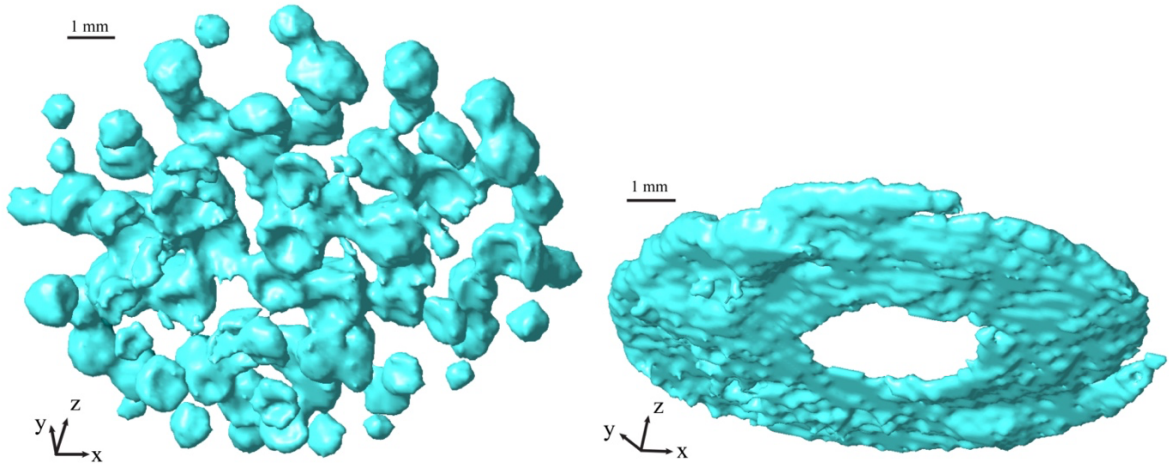

**Figure 7:** Segmented ice phases from reconstructed CT scans of vitrified CVS1 in isochoric (left) and isobaric-piston (right) configurations. While a single solid mass of ice forms near the top of the chamber in the isobaric-piston system, millimeter-scale globular ice formations are observed in the isochoric system. This morphology is reminiscent of bubbles formed at randomly located nucleation sites during vapor cavitation or outgassing.

Pre-conditioning the solution would be one way to mitigate outgassing or cavitation. The isochoric system employed in this study closely emulates Berthelot tube systems used to study liquids at negative pressure [30,34,35]. In these studies, rigid containers are filled with a liquid and cooled, and in the absence of pre-existing bubbles or cavitation, the fluid can be metastably tensioned [7]. In order to generate these negative pressures, however, the system must be pre-conditioned by initially heating the sealed container in order to pressurize the liquid and force into solution any stubborn vapor trapped in crevices – which can act as a nucleation seed. Since this pre-conditioning would certainly injure any contained biological material, we do not perform this step in our experiments, nor was it part of the protocol employed in [24]. Alternatively, one could pre-compress the solution, choose a solution that contracts less than the chamber, or both.

**Table 2:** Volume fractions (%) of ice for the CVS1 trials and of vapor for the 50 wt% DMSO trials. Chamber configurations are depicted in Figure 1. Representative CT reconstructions are shown in Figures 5 and 6.

| | Ice fraction - CVS1 | | Vapor fraction - 50 wt% DMSO | |
|---|---|---|---|---|
| | Isochoric | Isobaric (piston) | Isochoric | Isobaric (Kapton) |
| Trial 1 | 0.33 | 0.53 | 5.8 | 5.2 |
| Trial 2 | 0.44 | 0.75 | 5.5 | 8.9 |
| Trial 3 | 0.90 | 1.07 | 5.9 | 6.1 |
| Average | 0.56 | 0.78 | 5.7 | 6.7 |

## 4. Conclusions

Using a laboratory-scale photon counting X-ray CT system, we compare vitrification and ice formation processes in two aqueous solutions, CVS1 and 50 wt% DMSO, each under isochoric and isobaric conditions. We plunge cool these two solutions in liquid nitrogen (achieving cooling rates of roughly 110°C/min) and evaluate the presence and volume fraction of glass, ice, and vapor phases after equilibration. A free air-liquid interface may result in local cooling rates dipping below the rate necessary for vitrification and cause significant ice formation, muddying the effects of confinement. After introducing a movable aluminum piston to match the thermal surface conditions of the isochoric confinement, we find that isochoric confinement yields similar volumes of ice (for CVS1) and vapor (for 50 wt% DMSO) compared to the isobaric configuration. Despite the similar ice volume fractions of the two configurations, we observe that isochoric vitrification of CVS1 produces a unique ice morphology that could be the result of outgassing or cavitation. Either pre-compressing the solution, adding an expanding material (e.g. bismuth or antimony), choosing a solution whose thermal contraction is less than the chamber, or some combination thereof, is necessary to prevent outgassing or cavitation to maintain isochoric conditions during cooling.

Furthermore, our findings highlight the importance of aqueous solution and chamber thermo-mechanical properties for interpreting pressure measurements during vitrification in rigid isochoric systems. Without a better understanding of these properties, pressure measurements are best used as a quick tool for comparing formation of and relative volumes of ice formed so long as the ice volume is large enough to produce a pressure rise.

Additionally, the observation of unique ice morphologies motivates future investigation of ice growth during vitrification in isochoric systems. The presence of ice acts to self-pressurize the system and the rejection of solutes from the growing ice phase would increase the concentration of the unfrozen solution, further modulating its melting point and passively diffusing additional solutes into the biomaterial [15,27]. If the biospecimen is protected from an advancing ice front, the combined effects of higher pressure and increased solute concentration would increase the likelihood of successful vitrification in these portions of the systems.

Regarding the measurement system developed for this study, photon counting X-ray CT has proven to be a promising approach for the investigation of vitrification in aqueous systems. PCDs can produce a higher signal-to-noise ratio than traditional X-ray detectors, making them well suited for low contrast, high scatter systems such as isochoric thermodynamics. Moreover, traditional X-ray CT scans such as those done in [35]-[38] are not photon energy resolving, but photon counting X-ray CT can be. Due to this unique feature, photon counting X-ray CT could potentially identify different materials in a single CT scan by reconstructing each voxel's mass attenuation coefficient curves, identifying K-edges, or both. This capability could be used to quantitatively study processes such as the diffusion of CPAs into tissue and the formation of multiple thermodynamic phases simultaneously.

Lastly, the laboratory-scale system employed in this study images the chamber at a projection resolution of 98 μm in order to capture the full isochoric system in a single scan. More projections per rotation can improve the CT scan quality, although that was not necessary for this study. Opting for a smaller field of view would enable higher resolutions as the projection resolution scales linearly with magnification factor, M, and detector resolution. As an example, the current setup at M = 10 would yield a projected resolution of 17.2 μm. Volumes of ice at this scale are below the detectable limits possible with differential scanning calorimetry [16], as well as most visual [20] and X-ray diffraction [18] methods. Furthermore, assuming ideal conditions where phase segmentation is possible from single voxels, we can place a theoretical lower detection limit to volume fraction detection of $O(10^{-8})$ for this detector. Of course, realistically this number will be higher since more than one voxel is needed to confidently identify a phase distinction. Detectors with more pixels (and therefore more reconstructed voxels) would have lower limits. Photon counting X-ray CT can thus prove valuable for investigating further fundamental aspects of ice nucleation, growth, and vitrification.

### Acknowledgements

This work received financial support from the National Science Foundation (NSF) Engineering Research Center for Advanced Technologies for Preservation of Biological Systems (ATP-Bio) under NSF EEC Grant No. 194154. Anthony Consiglio received additional support from the NSF Graduate Research Fellowship under Grant No. DGE 1752814. Simo Mäkiharju and Jason Parker also gratefully acknowledge the support of NSF EAGER award #1922877 program managers Ron Joslin and Shahab Shojaei-Zadeh, which enabled the development of the photon counting CT system. We would also like to thank Dr. Angel Rodriguez for his instrumental contributions to building the X-ray imaging system.

## Declaration of competing interest

B.R. has filed a 2017 patent application related to isochoric vitrification, which is under review as of the date of submission of this work. B.R. also has financial interest in a commercial entity that holds the license to the said patent application. The other authors declare no competing interests.